# Coherence and uniqueness theorems for averaging processes in statistical mechanics

by

*Hugo H. Torriani*
*IMECC, UNICAMP, Caixa Postal 6065,*
*13081-970 Campinas, São Paulo, Brazil*
<torriani@ime.unicamp.br>

and

*Michiel Hazewinkel*
*CWI, POBox 94079,*
*1090GB Amsterdam, The Netherlands*
<mich@cwi.nl>

**Abstract**. Let $S$ be the set of scalings $\{n^{-1}: n=1,2,3,\cdots\}$ and let $L_z = z\mathbf{Z}^2$, $z \in S$ be the corresponding set of scaled lattices in $\mathbf{R}^2$. In this paper averaging operators are defined for plaquette functions on $L_z$ to plaquette functions on $L_{z'}$ for all $z', z \in S$, $z' = dz$, $d \in \{2,3,4,\cdots\}$ and their coherence is proved. This generalizes the averaging operators introduced by Balaban and Federbush. There are such coherent families of averaging operators for any dimension $D=1,2,3,\cdots$ and not only for $D=2$. Finally there are uniqueness theorems saying that in a sense, besides a form of straightforward averaging, the weights used are the only ones that give coherent families of averaging operators.

**MSCS**: 82B20.

**Key words and key phrases**: lattice theory, scaling, averaging operator, coarsening operator, scaling limit, field theory, coherent family of averaging operators, Balaban-Federbush averaging, plaquette function, renormalization, BF-average, coherent averaging.

## 1. Introduction.

Consider a family of lattices in 2-space, $L_z = z\mathbf{Z}^2 \subset \mathbf{R}^2 = \{(za, zb) \in \mathbf{R}^2: a, b \in \mathbf{Z}\}$, $z \in S$. The set $S$ is the set of length scales that is being discussed; for instance the set $S = \{2^{-r}: r = 1,2,3,\cdots\}$ as in [1, 2, 3, 5, 6, 7, 8], or $S = \{n^{-1}: n \in \mathbf{N}\}$, where $\mathbf{N}$ is the set of natural numbers $\mathbf{N} = \{1,2,3,\cdots\}$. These are the only two sets of length scales that will be used in this paper. Other sets of scalings can be used, such as $\mathbf{Q}_+$ the set of rational numbers greater than zero or $S = \{\prod_{p \in T} p^{a_p}: a_p \in \mathbf{Z}\}$ for $T$ a finite or infinite set of prime numbers.

Introduce a partial ordering on $S$ by $z \prec z'$ iff $z' = dz$, $d \in \{2,3,4,\cdots\}$. A partial ordering on a set $S$ is (downwards) directed if for all $z_1, z_2 \in S$ there is a $z \in S$ such that $z \prec z_1$, $z \prec z_2$. All the partially ordered sets mentioned so far are directed.

A plaquette of $L_z$ is a cell of $L_z$, that is a minimal square with corner points in $L_z$, that is, a square with with corner points $\{(az, bz), (az, (b+1)z), ((a+1)z, bz), ((a+1)z, (b+1)z)\}$ for some $(a,b) \in \mathbf{Z}^2$. Let $P(L_z)$ denote the set of plaquettes of $L_z$. A plaquette function is a function $f: P(L_z) \to \mathbf{R}$, or $\mathbf{C}$, or any other field of characteristic zero for that matter. Let $R(L_z)$ be the ring (vector space) of plaquette functions on $L_z$.

An averaging operator (also called coarsening operator) from scale $z$ to scale $z' = dz$, $d \in \mathbf{N}$ is a map $\alpha_d: R(L_z) \to R(L_{z'})$, $z' = dz$. One of the first conditions one requires of a collection of averaging operators for a set of length scales is coherence. That is, if $z'' = ez'$, $z' = dz$, then one should have

$$\alpha_e \circ \alpha_d = \alpha_{ed}. \tag{1.1}$$

In itself coherence is not all that difficult to achieve. For instance one can take straight averages or put the value of the averaged function on the larger plaquette equal to the value of the smaller plaquette situated at its lower left-hand corner, as illustrated in the two figures below for the case $d = 4$.

| 1/16 | 1/16 | 1/16 | 1/16 |
|------|------|------|------|
| 1/16 | 1/16 | 1/16 | 1/16 |
| 1/16 | 1/16 | 1/16 | 1/16 |
| 1/16 | 1/16 | 1/16 | 1/16 |

| 0 | 0 | 0 | 0 |
|---|---|---|---|
| 0 | 0 | 0 | 0 |
| 0 | 0 | 0 | 0 |
| 1 | 0 | 0 | 0 |

Here the large plaquette, bordered by heavy lines, is the union of 16 small plaquettes, and with obvious, though ad hoc notation

$$f^{\text{large plaquette}} = \frac{1}{16} \Big( \sum_{i,j=1,2,3,4} f_{i,j}^{\text{small plaquette}} \Big),$$

respectively,

$$f^{\text{large plaquette}} = f_{1,1}^{\text{small plaquette}}.$$

The coherent 'lower left-hand corner scheme' appears utterly daft; at least at the moment—in mathematics and mathematical physics one never knows what solutions to a given problem may one day turn out to be important.

There is also something quite counterintuitive about the straightforward averaging scheme. Intuitively the value of the plaquette function at a large plaquette is something like a field strength located at the center of that large plaquette. Thus it seems counterintuitive that it is made up of the field strengths of the smaller plaquettes without regard of how far the centers of these small plaquettes are removed from the center of the large plaquette; one would like to have some tapering off.

Far from unrelated to this intuitive reasoning is the following. Once one has a coherent scheme of averaging operators one has a (directed) inverse system (projective system) of vector spaces and vector space morphisms.

$$\alpha_d: R(L_z) \to R(L_{z'}); \quad z, z' \in S, \quad z' = dz. \tag{1.2}$$

See e.g. [4], chapter VIII. Then one can take the projective limit, which could be suggestively denoted with $R(L_0)$. An element of this projective limit is a family of functions $\{f^z \in R(L_z): z \in S\}$ such that $\alpha_d(f^z) = f^{z'}$ for all $z, z' \in S, \ z' = dz$. What one would like is some sort of decent relation between these projective limit elements and continuously differentiable functions on $\mathbf{R}^2$ so as to get some good relation between a coherent system of lattice models, indexed by a set of scales $S$, and a field theory. This does not happen for straightforward averaging but it does happen for Balaban-Federbush averaging in the sense that the continuous differentiable functions on $\mathbf{R}^2$ inject into the correponding projective limit. See [7] and the references quoted there.

For a picture of Balaban-Federbush averaging one positions the lattices involved differently, namely in such a way that the centers of the large plaquettes coincide with the centers of appropriate small plaquettes. In [1, 2, 3, 5, 6, 7, 8] the only averaging operators that occur are $\alpha_2$ and its iterates, and the picture for $\alpha_2$ is

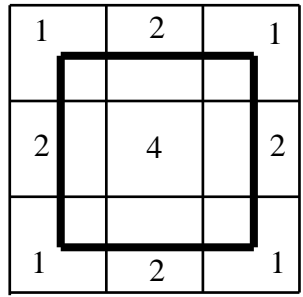

Here the large plaquette is bordered by heavy lines and the nine small plaquettes which affect the value of the averaged function on the large plaquette are bordered by thin lines. The numbers in the small plaquettes are their relative weights. They add up to 16 and so, using again obvious, but ad hoc and not very useful, notation, the formula is

$$f^{\text{large plaquette}} = 2^{-4}(f^{small}_{1,1} + 2f^{small}_{1,2} + f^{small}_{1,3} + 2f^{small}_{2,1} + 4f^{small}_{2,2} + 2f^{small}_{2,3} + f^{small}_{3,1} + 2f^{small}_{3,2} + f^{small}_{3,3}).$$

This particular rule is heuristically appealing in that the corner four small plaquettes influence precisely four large plaquettes, the four noncorner small edge plaquettes affect two large plaquettes and finally the center small plaquette only affects one large plaquette, suggesting that the relative weights should be exactly as they are in that in aggregate each small plaquette has exactly the same amount of total influence in the averaging process. This fails however for the iterates of the $\alpha_2$ such as $\alpha_2 \circ \alpha_2$.

It is also tempting to think that the right kind of averaging for scale changes that are powers of 2, or more generally any natural number larger than 1, would be:
    "To obtain the averaged value of $f^{\text{large}}$ at a given plaquette take a weighted sum of all the values of $f^{\text{small}}$ at those small plaquettes which intersect the large plaquette".
This fails for the iterate $\alpha_2 \circ \alpha_2$.

The right picture would appear to be as follows. Take a large plaquette $P$. Let $\tilde{P}$ be the plaquette with the same center and sides parallel to those of $P$ and of twice their length. Then the value of the averaged function at $P$ is a weighted sum of all small plaquettes completely contained in $\tilde{P}$. This is illustrated for the scale factors 2, 3, and 4 in the pictures below.

Balaban-Federbush averaging

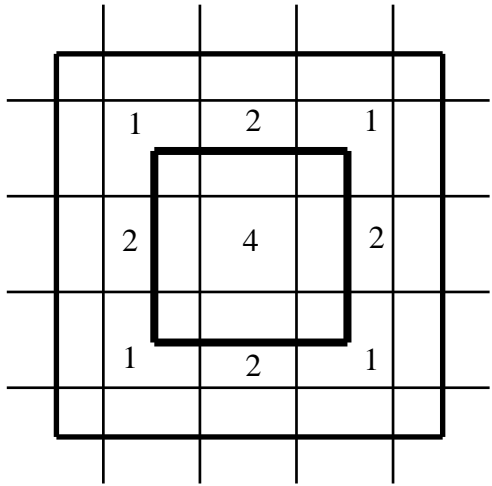
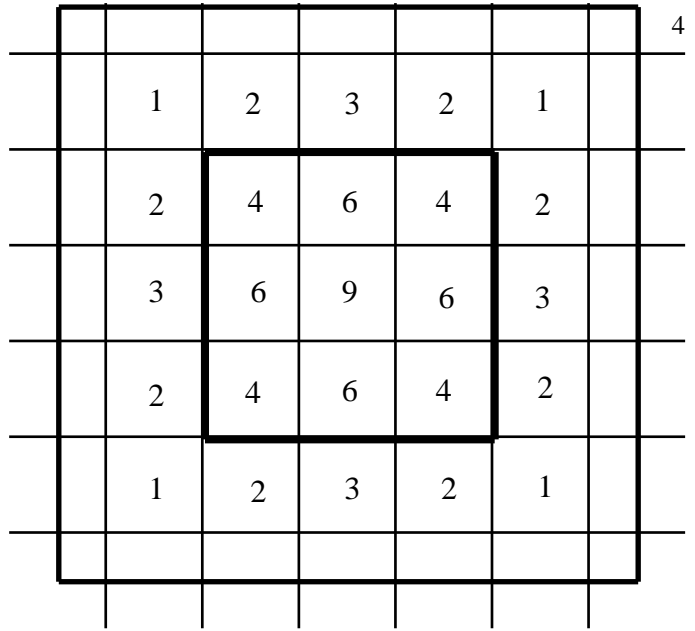
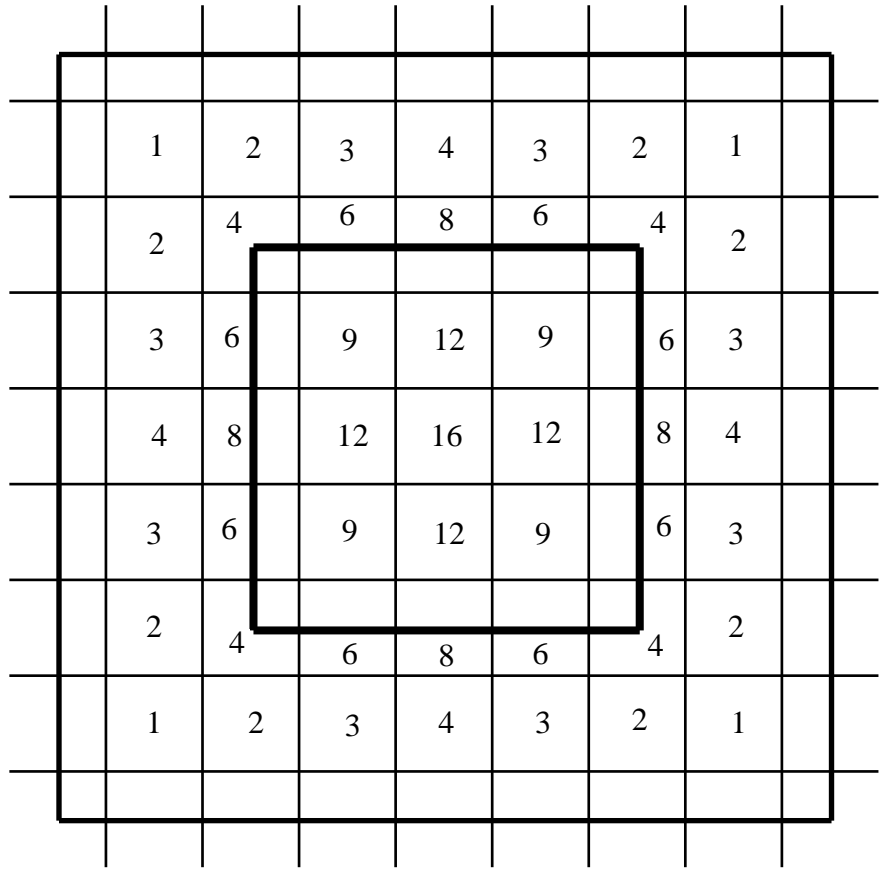

Here the small plaquettes are bordered with normal thickness lines; the large plaquette $P$ is bordered with heavy lines, and the plaquette $\tilde{P}$ that determines which small plaquettes influence the averaged value on the large plaquette is bordered by half heavy lines.

The relative weights are also indicated. For the scale factor $d$ there are precisely $(2d-1)^2$ small plaquettes which affect a large plaquette (as is easily checked). The relative weights in the pictures above add up to $16 = 2^4$, $81 = 3^4$, and $4^4$ respectively. So the true weights are respectively $2^{-4}$, $3^{-4}$, and $4^{-4}$ times the numbers indicated.

From these examples it is not difficult to guess what might be the general rule for any scale factor. And, as it happens, that obvious guess works to give a coherent system of averaging operators. Precise formulas will be given below in section 2. Moreover this scheme works not only in dimension $D = 2$, the plane case just discussed, but in any dimension $D = 1,2,3,4,\cdots$.

## 2. The averaging formula.

Consider a scale $z \in S$ and the corresponding lattice $L_z$. It is convenient to displace the coordinate system by $(\frac{1}{2}z, \frac{1}{2}z)$. Then the plaquettes of $L_z$ can be labelled by their centers wich have coordinates of the form $(az, bz)$, $a, b \in \mathbf{Z}$, see the figure below. Moreover the plaquettes of $L_{z'}$, $z' = dz$, $d \in \mathbf{N}$ have their centers at the points $(adz, bdz)$, $a, b \in \mathbf{Z}$.

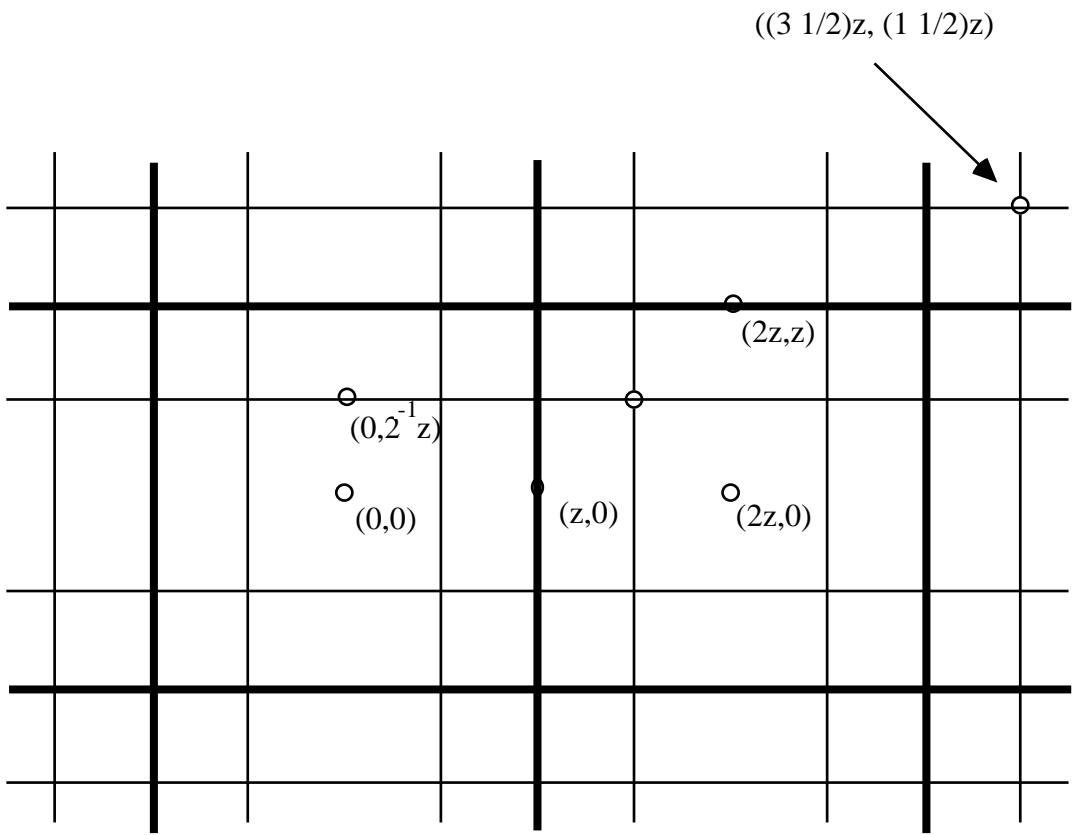

Let $f$ be a plaquette function on $L_z$ and denote its value on the plaquette with center $(az, bz)$ by $f_{az,bz}$. Let $z' = dz$. Then the BF-average $\alpha_d f$ is the plaquette function on $L_{z'}$ whose value at the plaquette $(adz, bdz)$ of $L_{z'}$ is given by

$$(\alpha_d f)_{adz,abz} = d^{-4} \sum_{|i|,|j| \leq d-1} (d - |i|)(d - |j|) f_{(da+i)z,(db+j)z} \quad . \tag{2.1}$$

Note that

$$\sum_{|i|,|j| \leq d-1} (d - |i|)(d - |j|) = d^4 ,$$

so that the weights in (2.1) add up to 1.

## 3. Proof of coherence.
The coherence theorem to be proved is now the following.

    3.1. *Theorem.* Let $S$ be the set of scales $S = \{2^{-r}: r = 0,1,2,3,\cdots\}$ or $S = \{n^{-1}: n \in \mathbf{N}\}$, and let the averaging operators $\alpha_d: R(L_z) \to R(L_{dz})$ be given by (2.1). Then

$$\alpha_e \circ \alpha_d = \alpha_{ed} \tag{3.2}$$

for all $d, e \in \mathbf{N}$.

    3.3. *Remark.* In case of the set of scales $S = \{2^{-r}: r = 0,1,2,3,\cdots\}$ there is a different way of looking at Theorem 3.1. Given any set of averaging operators

$$\alpha_2^z: R(L_z) \to R(L_{2z}), \ z, 2z \in S \ ,$$

define $\alpha_{2^r}^z: R(L_z) \to R(L_{2^r z})$ as the composite $\alpha_2^{2^{r-1}z} \circ \alpha_2^{2^{r-2}z} \circ \cdots \circ \alpha_2^z$. Then coherence is automatic (and one can even have the various $\alpha_2$ depend explicitly on $z$) because composition is associative. It is in this sense that the projective limits occurring in [7] are to be understood. The content of Theorem 3.1 in this case is that if $\alpha_2$ is given by (2.1) for $d = 2$, then its iterates are explicitly given by (2.1) for all $d = 2^r$. This is also the content of the main result of [9].

Proof of Theorem 3.1. By definition,

$$(\alpha_{ed} f)_{edaz, edbz} = (ed)^{-4} \sum_{|i|,|j| \leq ed-1} (ed - |i|)(ed - |j|) f_{(eda+i)z, (edb+j)z} \ . \tag{3.4}$$

On the other hand,

$$\begin{aligned}(\alpha_e(\alpha_d f))_{edaz, edbz} &= e^{-4} \sum_{|r|,|s| \leq e-1} (e - |r|)(e - |s|)(\alpha_d f)_{(eda+rd)z, (edb+sd)z} \\ &= e^{-4} d^{-4} \sum_{|r|,|s| \leq e-1} \sum_{|t|,|u| \leq d-1} (e - |r|)(e - |s|)(d - |t|)(d - |u|) f_{(eda+rd+t)z, (edb+sd+u)z} \ .\end{aligned} \tag{3.5}$$

So (3.4) will be equal to (3.5), proving (3.2) for all $f$, if and only if for all $|i|, |j| \leq ed - 1$ we have

$$\sum (e - |r|)(e - |s|)(d - |t|)(d - |u|) = (ed - |i|)(ed - |j|) \ , \tag{3.6}$$

where the sum on the left in (3.6) is over all solutions of the system of equalities and inequalities

$$rd + t = i, \quad |r| \leq e - 1, \ |t| \leq d - 1 \ , \tag{3.7}$$

$$sd + u = j, \quad |s| \leq e - 1, \ |u| \leq d - 1 \ . \tag{3.8}$$

The first step is to study the solutions of a system like (3.7).

    3.9. *Lemma.* Let $i$, $e$, $d$ be given with $i \leq ed - 1$ and consider the system of inequalites

and an equality (3.7). Then, depending on $i$, there are one or two solutions, as follows:
    (i) If $|i| > d(e-1)$ there is one solution, viz $r = e-1$, $t = i - d(e-1)$ if $i > 0$, and $r = -(e-1)$, $t = i + (e-1)d$ if $i < 0$.
    (ii) If $i$ is divisible by $d$, $i = kd$, there is precisely one solution, viz $r = k$, $t = 0$.
    (iii) If $|i| < d(e-1)$ and $i$ is not divisible by $d$, there are precisely two solutions, described as follows. Write $i = qd + p$, $p \in \{1, \cdots, d-1\}$. This can be done in a unique way. Then the two solutions are
$$r = q, \ t = p,$$
$$r = q+1, \ t = p - d.$$

The proof is routine.

    3.10. *Lemma.* For all $i$, $|i| \leq ed - 1$,

$$\sum_{\text{solutions of (3.7)}} (e-|r|)(d-|t|) \ = \ ed - |i|, \qquad (3.11)$$

where the sum on the left is, as indicated, over all pairs $(r,t)$ of integers such that (3.7) holds.

Proof. If $i > d(e-1)$, there is one solution, viz $r = e-1$, $t = i - d(e-1)$, and the left-hand side of (3.11) is equal to $(e - (e-1))(d - (i - (e-1)d)) = ed - i = ed - |i|$.
    If $i < -d(e-1)$, the only solution is $r = -(e-1)$, $t = i + (e-1)d$, and the left-hand side of (3.11) is equal to $(e - |-(e-1)|)(d - |i + (e-1)d|) = d - (-i - (e-1)d) = ed + i = ed - |i|$.
    If $i$ is divisible by $d$, $i = kd$, there is just one solution, viz $r = k$, $t = 0$ and the left-hand side of (3.11) is equal to $(e - |k|)d = ed - |kd| = ed - |i|$.
    Finally if $|i| < d(e-1)$ and $i$ is not divisible by $d$, there are precisely two solutions, viz $r = q$, $t = p$ and $r = q+1$, $t = p - d$. Note that for the first solution $t > 0$, and for the second $t < 0$. Thus for $q \geq 0$, so that also $i \geq 0$, the left-hand side of (3.11) is equal to

$$(e-q)(d-p) + (e-q-1)p = ed - qd - ep + qp + ep - qp - p = ed - qd - p = ed - |i|.$$

And for $q \leq -1$, so that $i \leq 0$, it is equal to

$$(e+q)(d-p) + (e+q+1)p = ed + qd + p = ed - |i|.$$

This proves Lemma 3.10.

Proof of Theorem 3.1 continued. Because the equations (3.7) and (3.8) are completely independent of each other, a solution of the combined system consists of picking a solution of one and combining it with a solution of the other. Thus, for given $i, j$, $|i|, |j| \leq ed - 1$, the sum on the left-hand side of (3.6) is equal to

$$\sum_{\text{Solutions of (3.7)}} (e-|r|)(d-|t|) \sum_{\text{Solutions of (3.8)}} (e-|s|)(d-|u|),$$

and this is equal to the right-hand side of (3.6) by Lemma 3.10.

## 4. Averaging in other dimensions than 2.
Now consider $D$-dimensional lattices $L_z = z\mathbf{Z}^D \subset \mathbf{R}^D$ and functions on the $D$-dimensional cells of $L_z$. Here $D$ is any natural number $1, 2, 3, \cdots$. Shift the coordinate system by the vector

$$(\underbrace{2^{-1}z, 2^{-1}z, \cdots, 2^{-1}z}_{D}).$$

Then the cells of $L_z$ are conveniently labelled by their center points which have coordinates of the form $(a_1z, a_2z, \cdots, a_Dz)$, $a_i \in \mathbf{Z}$, and the cells of $L_{z'}$, $z' = dz$ have center points with coordinates $(da_1z, da_2z, \cdots, da_Dz)$. The averaging formula is now

$$(\alpha_d f)_{da_1z, \cdots, da_Dz} = d^{-2D} \sum_{|i_1|, \cdots, |i_D| \leq d-1} (d - |i_1|) \cdots (d - |i_D|) f_{(da_1 + i_1)z, \cdots, (da_D + i_D)z} \ . \tag{4.1}$$

The proof that this is coherent is virtually identical with the proof given above for the case $D = 2$. The relevant identity to be proved is

$$\sum (e - |r_1|)(d - |t_1|) \cdots (e - |r_D|)(d - |t_D|) = (ed - |i_1|) \cdots (ed - |i_D|) \tag{4.2}$$

for all $|i_1|, \cdots, |i_D| \leq ed - 1$, where the left-hand sum is over all solutions of the system of equations and inequalities

$$\begin{aligned} i_1 &= r_1 d + t_1, \ |r_1| \leq e - 1, \ |t_1| \leq d - 1, \\ &\cdots \\ i_D &= r_D d + t_D, \ |r_D| \leq e - 1, \ |t_D| \leq d - 1 \ . \end{aligned} \tag{4.3}$$

Now, because the $D$ equation systems making up (4.3) are independent, a solution consists of picking a solution for each of the separate equation systems. So, for given $i_1, \cdots i_D$, $|i_1|, \cdots |i_D| \leq ed - 1$, the left hand side of (4.2) is equal to

$$\prod_{j=1}^{D} \sum_{\substack{\text{Solutions of the} \\ \text{j-th equation} \\ \text{of (4.3)}}} (e - |r_j|)(d - |t_j|) \ .$$

By Lemma 3.10 this is equal to the right-hand side of (4.2).

## 5. A uniqueness theorem.

One can consider averaging schemes like (4.1) in general with weights possibly different from the

$$d^{-2D}(d - |i_1|) \cdots (d - |i_D|) \tag{5.1}$$

of (3.2) and wonder for which weights this is coherent for the set of scales $S = \{n^{-1}: n \in \mathbf{N}\}$ for $D = 1$ (and hence for all $D$). The following uniqueness theorem says that, besides a straightforward averaging type scheme, the weights (5.1) are the only ones that work for such schemes.

  5.2. *Theorem.* Consider averaging schemes

$$(\alpha_d f)_{da_1z, \cdots, da_Dz} = \sum_{|i_1|, \cdots, |i_D| \leq d-1} w_{i_1}^d w_{i_2}^d \cdots w_{i_D}^d f_{(da_1 + i_1)z, \cdots, (da_D + i_D)z} \ , \tag{5.3}$$

where $w_i^d = w_{-i}^d$ and suppose that these are coherent for $D = 1$ (and hence for all $D$). Suppose moreover that $w_1^2 \neq 0$ (so that $\alpha_2$ is nontrivial) and that the weights are $\neq 0$ and add up to 1, as they should. Then there are two possibilities:
  (i) the weights $w_i^d$ are equal to those of (5.1); i.e.,

$$w_i^d = d^{-2}(d-|i|) \ . \tag{5.4}$$

(ii) the weights $w_i^d$ are as follows:

$$w_i^d = \begin{cases} 0 \text{ if } d-i \text{ is even,} \\ d^{-1} \text{ if } d-i \text{ is odd.} \end{cases} \tag{5.5}$$

5.5. *Remark.* The second solution is a kind of straightforward averaging.

Proof. First consider the case $d=2, d=3$. Than coherence says that

$$\alpha_2 \circ \alpha_3 = \alpha_3 \circ \alpha_2 = \alpha_6 \ , \tag{5.6}$$

and this implies certain identities between $w_1^2 = x$ and $w_1^3 = y_1$, $w_2^3 = y_2$. Specifically, consider the coefficients of the $f_{6a+1}$ in $\alpha_6 f$. On the one hand we have to look at all solutions of

$$1 = 2r+s, \ |r|<3, \ |s|<2 \ ,$$

which are $r=0, s=1; \ r=1, s=-1$ and give the coefficient

$$(1-2y_1-2y_2)x + y_1 x \ ,$$

and on the other hand at solutions of

$$1 = 3t+u, \ |t|<2, \ |u|<3 \ ,$$

which are $t=0, u=1$ and $t=1, u=-2$ and give the coefficient

$$(1-2x)y_1 + xy_2 \ .$$

Thus coherence implies that

$$x - y_1 x - 2y_2 x = y_1 - 2xy_1 + xy_2 \ . \tag{5.6}$$

Similarly, looking at the coefficients of the $f_{6a+i}$ in $\alpha_6 f$ for $i=2,3,4$, one finds the equations

$$y_1(1-2x) = (1-2x)y_2 + xy_1 \ , \tag{5.7}$$

$$y_1 x + y_2 x = x(1-2y_1-2y_2) \ , \tag{5.8}$$

$$y_2(1-2x) = xy_1 \ . \tag{5.9}$$

Substitute (5.9) in (5.7) to get $y_1(1-2x) = 2xy_1$, so that $x=\frac{1}{4}$, or $y_1=0$. In the latter case, by (5.7), $y_2=0$ or $x=\frac{1}{2}$. But if $y_1=y_2=0$, then also $x=0$ by (5.6) which is not the case by hypothesis. Thus there are just two possibilities for $x$, viz:

a) $x=\frac{1}{4}$. Then $y_1=2y_2$ by (5.9). Also $y_1+y_2=\frac{1}{3}$ by (5.8). In this case we have

$$x=\tfrac{1}{4}, \ y_1=\tfrac{2}{9}, \ y_2=\tfrac{1}{9} \ ,$$

in agreement with (5.4) for $d=2$ and $d=3$.

b) $x = \frac{1}{2}$. Then $y_1 = 0$, and by (5.8) $y_2 = \frac{1}{3}$, as is the case of (5.5) for $d = 2$ and $d = 3$.

Now let $d$ be any odd natural number and for convenience write

$$z_i = w_i^d, \quad i = 0, 1, 2, \cdots, d-1.$$

Consider

$$\alpha_2 \circ \alpha_d = \alpha_d \circ \alpha_2 = \alpha_{2d}$$

and look at the coefficients of the $f_{2da+i}$ in $\alpha_{2d} f$ for $i = 0, 1, \cdots, d-1$. First look at $i$'s of the form $i = 2d - (2k+1)$ for $k = 1, 2, \cdots, 2^{-1}(d-3)$. The only solution of

$$2d - 2k - 1 = rd + s, \quad |r| < 2, \quad |s| < d$$

is $r = 1$, $s = d - (2k+1)$, which gives the coefficient $xz_{d-2k-1}$. On the other hand, the solutions of

$$2d - 2k - 1 = 2t + u, \quad |t| < d, \quad |u| < 2$$

are $t = p - k, u = -1$ and $t = p - k - 1, u = 1$, which yield the coefficient $z_{d-k}x + z_{d-k-1}x$. Thus we find the equations

$$\begin{aligned} z_{d-3} &= z_{d-1} + z_{d-2}, \\ z_{d-5} &= z_{d-2} + z_{d-3}, \\ &\cdots \\ z_2 &= z_{2^{-1}(d+3)} + z_{2^{-1}(d+1)}. \end{aligned} \quad (5.10)$$

Now look at $i$'s of the form $i = 2d - 2k$, $k = 1, 2, \cdots 2^{-1}(d-1)$. There is just one solution of

$$2d - 2k = dr + s, \quad |r| < 2, \quad |s| < d,$$

viz $r = 1, s = d - 2k$, which gives the term $xz_{d-2k}$. There is also just one solution of

$$2d - 2k = 2t + u, \quad |t| < d, \quad |u| < 2,$$

viz $t = p - k, u = 0$, which yields a term $z_{p-k}(1-2x)$. Thus,

$$\begin{aligned} z_{d-1}(1-2x) &= xz_{d-2}, \\ z_{d-2}(1-2x) &= xz_{d-4}, \\ &\cdots \\ z_{p-2^{-1}(p-1)}(1-2x) &= xz_1. \end{aligned} \quad (5.11)$$

Finally, for $i = 1$ one finds the equation

$$(1-2x)z_1 + xz_{d-1} = (1 - 2z_1 - \cdots - 2z_{d-1})x + z_1 x. \quad (5.12)$$

Now suppose that $x = \frac{1}{4}$. Then (5.11) and (5.10) combine to give

$$z_{d-k} = k z_{d-1}.$$

Substitute this last equality in (5.12) to find $z_{d-1} = d^{-2}$, so that in this case we have the solution (5.4). In the second case, when $x = \frac{1}{2}$, equations (5.11) say that $z_1 = z_3 = \cdots = z_{d-2} = 0$, and then (5.10) gives $z_2 = z_4 = \cdots = z_{d-1}$. Substitute these relations in (5.12) and find $z_{2k} = d^{-1}$, $k = 1, 2, \cdots, 2^{-1}(d-1)$. Thus, in this case (5.5) applies.

It remains to show that if $x = \frac{1}{2}$ the only solution for the weights is as specified by (5.5) also for the $w_i^{2d}$. This can be done by a straightforward calculation of $\alpha_2 \circ \alpha_d = \alpha_{2d}$ or by proving that the averaging scheme given by (5.5) is coherent, which is also fairly direct. For example, consider the case that $e$ is even and $d$ is odd in the relation $\alpha_e \circ \alpha_d = \alpha_{ed}$. We have to look at the solutions of

$$i = dr + s, \quad |r| < e, \quad |s| < d . \tag{5.13}$$

If $i$ is even, then we must have either ($r$ is even and $s$ even) or ($r$ is odd and $s$ is odd). Each solution $(r,s)$ contributes a summand $w_r^e w_s^d$. If $r$ is even, $w_r^e = 0$ because $e$ is even, and if $s$ is odd, $w_s^d = 0$, because $d$ is odd. Thus we get a coefficient zero in this case which fits with $w_{\text{even}}^{ed} = 0$ (because $ed$ is even).

If $i$ is odd, then we must have either ($r$ is even and $s$ is odd) or ($r$ is odd and $s$ is even). If $i$ is such that $|i| < d(e-1)$ and not divisible by $d$ there are two solutions of (5.13), and for precisely one of them $r$ is odd and the corresponding $s$ is even. In this case one gets a contribution $w_r^e w_s^d = e^{-1} d^{-1}$ because $e - r$ is odd and $d - s$ is odd; the other solution gives a contribution zero because for that one $r$ is even. If $|i| > (e-1)d$, $r$ is either $e-1$ or $1-e$ which are both odd. As $s$ is even, the single solution in this case also gives a contribution $w_r^e w_s^d = e^{-1} d^{-1}$. Finally, if $i$ is divisible by $d$, $i = kd$, then $k$ must be odd and therefore the single solution gives the contribution $w_k^e w_0^d = e^{-1} d^{-1}$.

The other three cases are handled similarly.

### 6. Second uniqueness theorem.
There are more general uniqueness theorems than Theorem 5.2. Basically it is not needed to assume factorization of weights like in the previous section.

    6.1. *Theorem.* Let the averaging operators $\alpha_d$ in dimension $D = 2$ be given by

$$(\alpha_d f)_{daz, dbz} = \sum_{|i|, |j| \leq d-1} w_{i,j}^d f_{(da+i)z, (db+j)z} .$$

Suppose they are coherent, and assume that the weights $w_{i,j}$ satisfy the symmetry conditions $w_{i,j} = w_{-i,j} = w_{i,-j} = w_{j,i}$ and the genericity conditions $w_{0,0}^2, w_{0,1}^2, w_{1,1}^2 \neq 0$. Then

$$w_{i,j}^d = d^{-4}(d - |i|)(d - |j|) ,$$

which are the weights used before.

    6.2. *Conjecture.* There is no real doubt that the same theorem holds in dimensions $> 2$, and that the same proof will work (though it will become notationally a bit more complicated). That is, assume that the coherent averaging operators in dimension $D$ are given by

$$(\alpha_d f)_{da_1 z, da_2 z, \cdots, da_D z} = \sum_{|i_1|, \cdots |i_D| \leq d-1} w_{i_1, i_2, \cdots, i_D}^d f_{(da_1 + i_1)z, \cdots, (da_D + i_D)z} .$$

Assume the symmetry conditions

$$w^d_{i_{\sigma(1)},\cdots,i_{\sigma(D)}} = w^d_{i_1,\cdots,i_D}$$

for all permutations of $\{1,2,\cdots,D\}$, and

$$w^d_{i_1,\cdots,i_D} = w^d_{|i_1|,\cdots,|i_D|} \ .$$

Then,

$$w^d_{i_1,\cdots,i_D} = d^{-2D}(d-|i_1|)\cdots(d-|i_D|) \ .$$

Proof of Theorem 6.1.
As in the case of Theorem 5.2, first consider $\alpha_2(\alpha_3 f)) = \alpha_3(\alpha_2 f)$ and calculate the coefficients of $f_{(6a,6b)+(i,j)}$ in the two indicated ways.

First take $(i,j) = (0,4)$. This means we have to look at all solutions of

$$0 = 2r+s, \ 4 = 2t+u, \ |r|,|t| < 3, \ |s|,|u| < 2 \tag{6.3}$$

on the one hand, and at those of

$$0 = 3a+b, \ 4 = 3c+d, \ |a|,|c| < 2, \ |b|,|d| < 3 \tag{6.4}$$

on the other. The only solution of (6.3) is $r=0, s=0, t=2, u=0$, which gives the term

$$w^3_{0,2} w^2_{0,0} ,$$

and the only solution of (6.4) is $a=0, b=0, c=1, d=1$, which gives the term

$$w^2_{0,1} w^3_{0,1} \ .$$

Thus,

$$w^3_{0,2} w^2_{0,0} = w^2_{0,1} w^3_{0,1} \ . \tag{6.5}$$

Now look at $(i,j) = (0,2)$. This time the equations are

$$\begin{aligned}0 = 2r+s, \ 2 = 2t+u, \ |r|,|t| < 3, \ |s|,|u|<2, \\ 0 = 3a+b, \ 2 = 3c+d, \ |a|,|c| < 2, \ |b|,|d| < 3.\end{aligned} \tag{6.6}$$

The first one has the unique solution $r=0, s=0, t=1, u=0$, and the second one has two solutions: $a=0, b=0, c=1, d=-1$ and $a=0, b=0, c=0, d=2$. Hence

$$w^3_{0,1} w^2_{0,0} = w^2_{0,1} w^3_{0,1} + w^2_{0,0} w^3_{0,2} \ . \tag{6.7}$$

Combining this with (6.5) and using $w^2_{0,0} \neq 0$ one finds

$$w^3_{0,1} = 2 w^3_{0,2} \ . \tag{6.8}$$

Now consider $(i,j) = (0,1)$. This gives

$$w_{0,1}^3 w_{0,1}^2 + w_{0,0}^3 w_{0,1}^2 = w_{0,1}^2 w_{0,2}^3 + w_{0,0}^2 w_{0,1}^3 ,$$

and combining this with (6.5), (6.8), and using $w_{0,1}^2 \neq 0$, there results

$$w_{0,0}^3 = 3 w_{0,2}^3 . \qquad (6.9)$$

Now look at $(i,j) = (3,3)$ to find (using $w_{1,1}^2 \neq 0$)

$$w_{1,1}^3 + 2 w_{1,2}^3 + w_{2,2}^3 = w_{0,0}^3 . \qquad (6.10)$$

Now, also

$$1 = w_{0,0}^3 + 4 w_{0,1}^3 + 4 w_{0,2}^3 + 4 w_{1,1}^3 + 8 w_{1,2}^3 + 4 w_{2,2}^3 .$$

Combining this with (6.10), (6.8), (6.9) there results

$$w_{0,2}^3 = \tfrac{1}{27}, \ w_{0,1}^3 = \tfrac{2}{27}, \ w_{0,0}^3 = \tfrac{1}{9} , \qquad (6.11)$$

and from (6.5),

$$w_{0,0}^2 = 2 w_{0,1}^2 . \qquad (6.12)$$

Next look at $(i,j) = (4,4)$ and $(i,j) = (4,5)$. This gives

$$w_{2,2}^3 w_{0,0}^2 = w_{1,1}^2 w_{1,1}^3, \ w_{2,2}^3 w_{0,1}^2 = w_{1,1}^2 w_{1,2}^3 . \qquad (6.13)$$

Using (6.12), these relations give $w_{1,1}^2 w_{1,1}^3 = w_{2,2}^3 w_{0,0}^2 = 2 w_{2,2}^3 w_{0,1}^2 = 2 w_{1,1}^2 w_{1,2}^3$, whence

$$w_{1,1}^3 = 2 w_{1,2}^3 . \qquad (6.14)$$

Next look at $(i,j) = (3,5)$ to find

$$w_{1,2}^3 + w_{2,2}^3 = w_{0,2}^3 , \qquad (6.15)$$

and combine this with (6.14), (6.11), (6.10), to find the remaining values of the $w_{i,j}^3$, viz

$$w_{1,1}^3 = \tfrac{4}{81}, \ w_{1,2}^3 = \tfrac{2}{81}, \ w_{2,2}^3 = \tfrac{1}{81} . \qquad (6.16)$$

Now put this in (6.13) and use $w_{0,0}^2 + 4 w_{0,1}^2 + 4 w_{1,1}^2 = 1$. This gives

$$w_{1,1}^2 = \tfrac{1}{16}, \ w_{0,1}^2 = \tfrac{1}{8}, \ w_{0,0}^2 = \tfrac{1}{4} . \qquad (6.19)$$

Thus the $w_{i,j}^d$ for $d = 2$, and $d = 3$ have the right values.

By the coherence assumption and Theorem 3.1, it now suffices to prove that the $w_{i,j}^p$ have the stated values for $p$ an odd prime. Actually the following arguments work for any odd natural number $>1$ and similar arguments can be given for even numbers.

Consider the coefficients of the $f_{(2pa, 2pb)+(i,j)}$ in $\alpha_{2p} f = \alpha_2 (\alpha_p) = \alpha_p (\alpha_2 f)$ calculated in the two ways indicated.

First consider pairs $(i,j)$ of the form $(2p-2k, 2p-2l-1)$, $k = 1, 2, \cdots, \tfrac{p-1}{2}, l = 0, 1, \cdots, \tfrac{p-1}{2}$.

This gives $w^p_{p-k,p-l}w^2_{0,1} + w^p_{p-k,p-l-1}w^2_{0,1} = w^2_{1,1}w^p_{p-2k,p-2l-1}$, so that

$$2w^p_{p-k,p-l} + 2w^p_{p-k,p-l-1} = w^p_{p-2k,p-2l-1}. \tag{6.20}$$

Here, for economy of notation, $w^p_{i,j} = 0$ if $i > p-1$ or $j > p-1$. Taking $l = 0, k = 1$ in (6.20) we see that

$$w^p_{p-2,p-1} = 2w^p_{p-1,p-1}. \tag{6.21}$$

Now consider pairs of the form $(i,j) = (2p-2k, 2p-2l)$, $k,l = 1,2,\cdots,\frac{p-1}{2}$, to find

$$w^p_{p-2k,p-2l} = 4w^p_{p-k,p-l}, \tag{6.22}$$

and in particular,

$$w^p_{p-2,p-2} = 4w^p_{p-1,p-1}. \tag{6.23}$$

Finally, consider pairs of the form $(i,j) = (2p-2k-1, 2p-2l-1)$, $k,l = 0,1,\cdots,\frac{p-1}{2}$, to find

$$w^p_{p-k,p-l} + w^p_{p-k,p-l-1} + w^p_{p-k-1,p-l} + w^p_{p-k-1,p-l-1} = w^p_{p-2k-1,p-2l-1}. \tag{6.24}$$

With induction, starting from (6.21) and (6.23), and using (6.20), (6.22), (6.24), as the case may be, it follows that

$$w^p_{p-i,p-j} = ij w^p_{p-1,p-1}, \quad i,j = 1,2,\cdots,p-1. \tag{6.25}$$

Further,

$$\sum_{i,j} w^p_{i,j} = p^4, \tag{6.26}$$

which combined with (6.25) gives

$$w^p_{p-1,p-1} = p^{-4}. \tag{6.27}$$

This shows that the weights $w^p_{i,j}$ have the required values and finishes the proof of Theorem 6.1.

6.28. *Remark.* It is natural to take for the 'weights' elements of the same field in which the plaquette functions take their values. For instance in the case of complex valued plaquette functions the weights in the statement of Theorem 6.2 can be complex numbers.

However, in that case, it might in some settings be natural to take another normalizing condition than that the weights sum to 1, viz, that they form a complex vector of norm 1, i.e. $(\sum_{i,j} \|w^d_{i,j}\|^2)^{\frac{1}{2}} = 1$. For this normalization there are other solutions. They are all as follows. For each prime number $p$ there is a complex number $\zeta_p$ of norm 1. For each natural number $d = 2,3,\cdots$ write it as a product of prime numbers, $d = p_1^{a_1} \cdots p_r^{a_r}$. Then,

$$w^d_{i,j} = d^{-4}(\zeta_{p_1}^{a_1} \cdots \zeta_{p_r}^{a_r})(d-|i|)(d-|j|).$$

### 7. Third uniqueness theorem.

Now let us consider again lattices $L_z = z\mathbf{Z}^2 \subset \mathbf{R}^2 = \{(za, zb) \in \mathbf{R}^2 : a, b \in \mathbf{Z}\}$, $z \in S$, where $S$ is a set of length scales and, as in the first part of the introduction, consider averaging operators which are of the form:

"Value of the averaged plaquette function on a large plaquette is a weighted sum of the values of the plaquette function being averaged on the small plaquettes contained in that large plaquette."

Label plaquettes by the coordinates of their lower left-hand corner. Then the general formula is

$$(\alpha_d f)_{daz, dbz} = \sum_{0 \leq i, j \leq d-1} w^d_{i,j} f_{(da+i)z, (db+j)z} . \tag{7.1}$$

Here the $w^d_{i,j}$, $0 \leq i, j \leq d-1$, are a set of nonnegative numbers that add up to one.

As before, when one is working with a set of scales of the form $S = \{2^{-r} : r = 0, 1, 2, \cdots\}$, or, more generally, $S = \{d^{-r} : r = 0, 1, 2, \cdots\}$, $d$ any fixed natural number $\geq 2$, one can choose $\alpha_2$, resp. $\alpha_d$, arbitrarily and define the $\alpha_{d^s}$ as the $s$-fold iterates of $\alpha_d$. There results a quite simple formula for these iterates. Indeed,

$$(\alpha_{d^s} f)_{d^s az, d^s bz} = \sum_{0 \leq i, j \leq d^s - 1} w^{d^s}_{i,j} f_{(d^s a + i)z, (d^s b + j)z} , \tag{7.2}$$

$$w^{d^s}_{i,j} = w^d_{i_1, j_1} w^d_{i_2, j_2} \cdots w^d_{i_s, j_s} , \tag{7.3}$$

with

$$i = i_1 d^{s-1} + i_2 d^{s-2} + \cdots + i_{s-1} d + i_s, \ 0 \leq i_1, i_2, \cdots, i_s \leq d-1$$

and

$$j = j_1 d^{s-1} + j_2 d^{s-2} + \cdots + j_{s-1} d + j_s, \ 0 \leq j_1, j_2, \cdots, j_s \leq d-1$$

being the $d$-adic expansions of $i$ and $j$, i.e. the unique ways of writing $i$ and $j$ in the forms indicated.

Now, let us return to the case of the full set of scales $S = \{n^{-1} : n = 1, 2, \cdots\}$. There are a number of fairly obvious coherent sets of averaging processes. For instance, the straightforward averaging scheme

$$w^d_{i,j} = d^{-2}, \ i, j = 0, 1, \cdots, d-1 ; \tag{7.4}$$

the four corner schemes

$$w^d_{0,0} = 1, \text{ all other weights zero}; \quad w^d_{0, d-1} = 0, \text{ all other weights zero};$$
$$w^d_{d-1, 0} = 1 \text{ all other weights zero}; \quad w^d_{d-1, d-1} = 1, \text{ all other weights zero};$$

the diagonal scheme

$$w^d_{i,i} = d^{-1}, \ i = 0, 1, \cdots, d-1, \text{ all other weights zero};$$

and the antidiagonal scheme

$$w^d_{i,d-1-i} = d^{-1}, \ i = 0,1,\cdots,d-1, \text{ all other weights zero.}$$

For the case of only odd scale length changes, i.e. $S = \{(2n+1)^{-1}: n = 0,1,2,\cdots\}$, there is in addition the central scheme

$$w^d_{\frac{d-1}{2},\frac{d-1}{2}} = 1, \text{ all other weights zero,}$$

which also has a coherent analogue in the case of Balaban-Federbush type averaging.

All of these, except (7.4), are degenerate in some sense, and by the theorem below, under very mild genericity (= nondegeneracy) conditions, straightforward averaging according to (7.4) is the only coherent family of averaging operators.

7.5. *Theorem.* Consider averaging processes of the type (7.1), with the set of length scales $S = \{n^{-1}: n = 1,2,\cdots\}$ and weights $w^d_{i,j}$, $i,j = 0,1,\cdots,d-1$, that sum to 1. Assume moreover that in case the scale changes by a factor of 2 the four weights are nonzero, i.e. $w^2_{00} \neq 0$, $w^2_{0,1} \neq 0$, $w^2_{1,0} \neq 0$ and $w^2_{1,1} \neq 0$. Then $w^d_{i,j} = d^{-2}$, $0 \leq i,j \leq d-1$.

Proof. For coherence we need again of course $\alpha_e(\alpha_d(f)) = \alpha_{ed}(f)$ for all natural numbers $d$ and $e$. Put in (7.1). This works out as the condition that the weights must satisfy

$$w^e_{r,t} w^d_{s,u} = w^{ed}_{i,j} \quad \text{for all } 0 \leq i,j \leq ed-1, \tag{7.6}$$

where $r,s,t,u$ are uniquely determined by

$$\begin{aligned} rd + s = i, \ 0 \leq r \leq e-1, \ 0 \leq s \leq d-1, \\ td + u = j, \ 0 \leq t \leq e-1, \ 0 \leq u \leq d-1 . \end{aligned} \tag{7.7}$$

Note that the system of equalities and inequalities (7.7) always has precisely one solution. This simplifies things considerably. As in the proofs of the two previous uniqueness theorems, first consider $e = 3$, $d = 2$ and use the condition $\alpha_3 \circ \alpha_2 = \alpha_2 \circ \alpha_3$. Take for instance $(i,j) = (2,3)$. Then, on the one hand, we must look at the equations

$$\begin{aligned} 2r + s = 2, \ 0 \leq r \leq 2, \ 0 \leq s \leq 1, \\ 2t + u = 3, \ 0 \leq t \leq 2, \ 0 \leq u \leq 1, \end{aligned}$$

with unique solution $r = 1$, $s = 0$, $t = 1$, $u = 1$, which gives the term $w^3_{1,1} w^2_{0,1}$; and, on the other, at the equations

$$\begin{aligned} 3k + l = 2, \ 0 \leq k \leq 1, \ 0 \leq l \leq 2, \\ 3m + n = 3, \ 0 \leq m \leq 1, \ 0 \leq n \leq 2, \end{aligned}$$

with unique solution $k = 0$, $l = 2$, $m = 1$, $n = 0$ which gives the term $w^2_{0,1} w^3_{2,0}$. These two terms must be equal, and so, using the genericity assumption that $w^2_{0,1} \neq 0$, one finds

$$w^3_{1,1} = w^3_{2,0} . \tag{7.8}$$

Similarly, using the pairs of indices $(0,2)$, $(0,3)$, $(2,0)$, $(2,2)$, $(2,5)$, $(3,3)$, $(5,2)$, one finds seven more equalities relations among the $w^3_{i,j}$. Together with (7.8), these suffice to prove all the $w^3_{i,j}$, $0 \leq i,j \leq 2$ equal, so they must all be equal to $\tfrac{1}{9}$. Using a few other $(i,j)$ pairs, e.g. $(0,1)$, (1,1), (1,3), this in turn gives $w_{0,0}^2 = w_{0,1}^2 = w_{1,0}^2 = w_{1,1}^2 = \frac{1}{4}$.

Now let $d$ be any natural number $\geq 4$ and consider $\alpha_d \circ \alpha_2 = \alpha_2 \circ \alpha_d$. First look at pairs of indices of the form $(i,j) = (2x, 2y)$, $0 \leq 2x, 2y \leq d-1$. The equations and inequalites to be considered at are

$$2r + s = i, \ 2t + u = j, \ 0 \leq r, t \leq d-1, \ 0 \leq s, u \leq 1 \ .$$

The solution is $r = x$, $t = y$, $s = 0$, $u = 0$, which gives the term $w_{x,y}^d w_{0,0}^2$. On the other hand, we must look at

$$i = dk + l, \ j = dm + n, \ 0 \leq k, m \leq 1, \ 0 \leq l, n \leq d-1 \ ,$$

with the unique solution $k = m = 0$, $l = 2x$, $n = 2y$, which gives the term $w_{0,0}^2 w_{2x,2y}^d$. (One uses $0 \leq 2x, 2y \leq d-1$.) So, equality of these two terms gives

$$w_{x,y}^d = w_{2x,2y}^d \quad \text{for} \quad 0 \leq 2x, 2y \leq d-1 \ . \tag{7.9}$$

Similarly, looking at pairs of indices of the form $(2x, 2y+1)$, $(2x+1, 2y)$, $(2x+1, 2y+1)$, in the appropriate ranges, one finds

$$\begin{aligned} w_{x,y}^d &= w_{2x,2y+1}^d \quad \text{for} \quad 0 \leq 2x, 2y+1 \leq d-1 \ , \\ w_{x,y}^d &= w_{2x+1,2y}^d \quad \text{for} \quad 0 \leq 2x+1, 2y \leq d-1 \ , \\ w_{x,y}^d &= w_{2x+1,2y+1}^d \quad \text{for} \quad 0 \leq 2x+1, 2y+1 \leq d-1 \ . \end{aligned} \tag{7.10}$$

With induction on $(i,j)$ the four equations suffice to prove that all the $w_{i,j}^d$, $0 \leq i, j \leq d-1$, are equal, and complete the proof of Theorem 7.5.

7.11. *Remark.* Note that no symmetry conditions are needed for Theorem 7.5.

7.12. *Remarks.* There are quite likely many more uniqueness results on coherent averaging schemes that can be proved. For instance one can wonder about coherent BF-type averaging schemes of the type

$$(\alpha_d f)_{daz, dbz} = \sum_{|i|, |j| \leq kd-1} w_{i,j}^d f_{(da+i)z, (db+j)z}$$

for a given natural number $k$ possibly greater than 1, so that the value of the averaged function on a large plaquette is influenced by a greater range of small plaquettes (than in the BF case). There are probably close connections between rates of falling off in such schemes and good relations between differentiable functions and elements of the projective limit corresponding to an averaging scheme.

**Acknowledgement.**
This paper would never have come into existence were it not for the preprint [9], which in a rather different and lengthier and more complicated way proves coherence for the case $D = 2$ and the set of length scales $\{2^{-r}: r \in \mathbf{N}\}$.